\documentclass[prb,twocolumn,showpacs]{revtex4}

\usepackage{amsmath}
\usepackage{bm}% bold math
\usepackage{graphicx}% Include figure files

\def\ket#1{|#1\rangle }
\def\bra#1{\langle#1 | }

\def\punkt{\;\; .}
\def\komma{\;\; ,}
\def\expect#1{\langle#1 \rangle}

\def\w{\omega}
\def\H{{\cal H}}
\def\e{\epsilon}

\def\Tr#1{\textrm{Tr}\left[#1\right]}
\def\non{\nonumber\\ }

\begin{document}

\title{A  Numerical Renormalization Group approach to Green's Functions for
Quantum Impurity Models}

\author{Robert Peters}
\affiliation{Institut f\"ur Theoretische Physik,
            Universit\"at G\"ottingen, D-37077 G\"ottingen, Germany}

\author{Thomas Pruschke}
\affiliation{Institut f\"ur Theoretische Physik,
             Universit\"at G\"ottingen, D-37077 G\"ottingen, Germany}

\author{Frithjof B. Anders}
\affiliation{Institut f\"ur Theoretische Physik, Universit\"at Bremen,
                  P.O. Box 330 440, D-28334 Bremen, Germany}

%%\date{30-06-2004}
%%\date{\today}
\date{07-19-2006}

\begin{abstract}
We present a novel technique for the calculation of dynamical
correlation functions of quantum impurity systems in equilibrium with
Wilson's numerical renormalization group. Our formulation is based on
a complete basis set of the Wilson chain. In contrast to all previous
methods, it does not suffer from overcounting of excitation. By
construction, it always fulfills sum rules for spectral
functions. Furthermore, it accurately reproduces local thermodynamic
expectation values, such as occupancy and magnetization, obtained
directly from the numerical renormalization group calculations.
\end{abstract}

\maketitle

\section{Introduction}

The understanding of quantum impurity systems
% in all parameter
% regimes 
became one the cornerstones of condensed matter theory during the
last decade. Quantum impurity systems
%regained interest in the last decade due to their 
appear at the heart of a variety of different physical problems. Traditionally,
they were used to describe the interaction of magnetic impurities with a
metallic host.\cite{Hewson93}  Nowadays, quantum impurity systems play a
fundamental role in our understanding of the low temperature properties
of single-electron transistors\cite{KastnerSET1992,NatureGoldhaberGordon1998}
and the tunneling spectroscopy of adatoms on
metal surfaces.\cite{Manoharan2000,AgamSchiller2000} 
The basic structure common to all such systems is a mesoscopic subsystem or 
device -- such as a quantum dot, an organic donor-acceptor molecule or an adatom -- coupled to a continuum of
states that can be represented by non-interacting particles of
either fermionic or bosonic nature. Typical realizations are models 
like the single impurity Anderson model,\cite{Anderson61} where the continuum
of states is described by free fermions, or the spin-boson model,\cite{Leggett1987}
where the discrete orbitals interact with a bosonic bath. In addition,
within the dynamical mean-field theory\cite{Pruschke95,Georges96} or
its cluster extensions\cite{MaierJarrellPruschkeHettler2005} lattice
models for strongly correlated fermions have been mapped onto quantum
impurity problems embedded in a fictitious, self-consistent
bath. Obtaining a self-consistent solution for these theories requires
very accurate determination of local Green's functions. 

Wilson's numerical renormalization group (NRG) approach\cite{Wilson75} to 
quantum impurity problems is one of the most
powerful and flexible ways for accurately calculating thermodynamic
properties of quantum impurity systems. In addition, it provides a
deep insight into the underlying physics through the analysis of the fixed
point Hamiltonians.\cite{KrishWilWilson80a,KrishWilWilson80b,CoxZawa98} The key
ingredient is a logarithmic discretization of the bath continuum,
resulting in a well defined hierarchy of energy or temperature scales. 
The discretized model is then iteratively diagonalized and the basis
set truncated, retaining only those states with low lying energies after
each step.  Each iteration represents a certain temperature $T$, and 
all  thermodynamic properties are determined for that particular
$T$.\cite{Wilson75,KrishWilWilson80a} For calculating
dynamical properties\cite{CostiHewsonZlatic94,Costi97,BullaHewsonPruschke98,Hofstetter2000,BullaCostiVollhardt01} three fundamental problems arise:
(i) how to recover the continuum limit from a discretized spectrum, (ii) 
how to obtain dynamical information on all energies scale at
arbitrarily low temperature in such away that (iii) spectral sum
rules are always fulfilled and thermodynamics expectation
values are reproduced {\em exactly} independent of how many states are
kept after each NRG step. 

In this paper, we will derive an exact analytical expression for arbitrary
dynamical correlation functions which solves this problems. It
is based upon the recent identification of a {\em complete basis set} of
the Wilson chain, which is also an approximate eigenbasis of the NRG
Hamiltonian.\cite{AndersSchiller2005, AndersSchiller2006} We will show
that this complete basis set 
automatically ensures that spectral sum rules are fulfilled exactly and
thermodynamic expectation values reproduced accurately by the spectral
functions. We furthermore explicitly demonstrate that our novel approach yields
spectral functions which are largely insensitive to the number of eigenstates
kept in each NRG iteration, thus improving the applicability of the
NRG to multi-orbital and multi-site impurity problems and consequently
also to multi-band and cluster DMFT problems, because the number of
NRG states needed can be significantly reduced without loosing
accuracy. In addition, phenomenological patching algorithms typically
employed to combine excitations from different NRG
iterations\cite{BullaCostiVollhardt01} become obsolete, since we
rigorously identify which excitations actually contribute at which
Wilson shell.

The paper is organized as follows. In the next section we will briefly
review the basic theoretical concepts and 
derive the complete basis set. We furthermore show how this complete basis
set can be used to calculate dynamical correlation functions.
Section III is devoted to a detailed discussion of results for
the simplest and most important quantum impurity model, the single impurity
Anderson model. We will explicitly compare our new method to the standard
implementations. A summary and outlook in section IV will conclude the paper.
Proofs for our claim that sum rules and spectral averages are
automatically respected within this formulation are given in the appendix.

\section{Theory}
\label{sec:theory}

The NRG is a very powerful tool for accurately calculating
equilibrium properties of 
%arbitrarily complex <-- ein wenig uebertrieben, oder?
%% n�, finde ich nicht. 
quantum impurity models.
Originally developed for treating the single-channel, single-impurity Kondo
Hamiltonian,\cite{Kondo62,Wilson75} this non-perturbative approach
was successfully extended to the Anderson impurity
model,\cite{Anderson61,KrishWilWilson80a,KrishWilWilson80b} the
two-channel Anderson\cite{AndersTCSIAM2005} and Kondo 
Hamiltonian,~\cite{Cragg_et_al,PangCox91} different two-impurity
clusters,~\cite{Jones_et_al_1987,Jones_et_al_1988,Sakai_et_al_1990,
Sakai_et_al_1992a,Sakai_et_al_1992b,IJW92,pruschkeDCA04} and a host of related
zero-dimensional problems. Recently, it was extended to 
equilibrium properties of  impurity models with a {\em bosonic} bath 
\cite{BullaBoson2003,BullaVoita2005}
or even combinations of both fermionic and Bosonic
baths.\cite{GlossopIngersent2005}

The Hamiltonian of a quantum impurity system is generally
given by
\begin{eqnarray}
{\cal H} = {\cal H}_{\rm bath} + {\cal H}_{\rm imp} +
           {\cal H}_{\rm mix} \; ,
\end{eqnarray}
where ${\cal H}_{\rm bath}$ models the continuous bath,
${\cal H}_{\rm imp}$ represents the decoupled impurity, and
${\cal H}_{\rm mix}$ describes the coupling between the two
subsystems. Thermodynamic properties of such a quantum impurity system
are very accurately obtained using the NRG. At the heart of this approach is
a logarithmic discretization of the continuous bath, controlled
by the discretization parameter $\Lambda > 1$;~\cite{Wilson75}
the continuum limit is recovered for $\Lambda \to 1$. Using
an appropriate unitary transformation,\cite{Wilson75} the
Hamiltonian is mapped onto a semi-infinite chain, with the
impurity coupled to the open end. By construction, the $N^{\rm th}$
site of this chain couples only to its immediate neighbors, which
allows to write the Hamiltonian of the infinite chain as limit for
$N\to\infty$ of a sequence of finite chains, denoted by $H_N$, with a
unique prescription $H_N\mapsto H_{N+1}$, the RG equation of the
NRG. The $N^{\rm th}$ link along the 
chain represents an exponentially decreasing energy scale
$D_N \sim \Lambda^{-N/2}$ for a fermionic\cite{Wilson75}
and $D_N \sim \Lambda^{-N}$ for a bosonic
bath.\cite{BullaBoson2003} Using this hierarchy of scales,
the sequence of finite-size Hamiltonians ${\cal H}_N$ for the
$N$-site chain\footnote{We use the standard notation, by
which the $N$-site chain contains the impurity degrees of
freedom, as well as the first $N + 1$ Wilson shells (labelled
by $n = 0, \cdots, N$.)}
is solved iteratively, i.e.\ starting with $N=0$ one constructs the
Hamiltonian $H_0$, diagonalizes it, adds the next site to obtain
$H_1$, diagonalizes it etc. The problem of an exponentially increasing
Hilbert space is resolved by observing, that, due to the exponential
decrease of the energy scales, only low-energy states actually
contribute in the step $N\to N+1$; one thus discards the high-energy
states before moving on to the next step to maintain a manageable number
of states, we denote with $N_S$ in the following. The reduced basis set of ${\cal H}_N$ obtained that way
is expected to faithfully describe the spectrum of the full
Hamiltonian on a scale of the order of $D_N$, corresponding to a
temperature $T_N \sim D_N$.\cite{Wilson75} Note that even from this
brief discussion it is evident that for a given step $N$ all
information about energies $E\gg D_N$ has been lost, while no
information about energy scales $E \ll D_N$ is available yet. Thus, to
calculate dynamical quantities with a similar accuracy as
thermodynamic, one has to tackle the problem of correctly mixing information
from earlier and later NRG steps.

\subsection{Complete Basis Set }

In a recent extension of the NRG to real-time dynamics out of
equilibrium\cite{AndersSchiller2005,AndersSchiller2006} a {\em
 complete basis set } for such  a Wilson chain of length $N$ has been
identified. Furthermore, this complete basis set also forms an
approximate eigenbasis of the Hamiltonian $H_{N}$. Since this complete basis
set plays a crucial role in the derivation of the analytical
expression for spectral functions,  we summarize briefly the main
ideas\cite{AndersSchiller2005} of the proof of completeness discussed
extensively in Ref.~\onlinecite{AndersSchiller2006}.

At first sight  the  claim that the NRG automatically generates a complete
basis set, which is also an approximate eigenbasis of the chain
Hamiltonian $H_N$, might appear contradictory. A renormalization
procedure is usually viewed as a clever way to identify the relevant
degrees of freedom by reducing the dimensionally of the underlying
Fock space. In order to excavate the optimally adapted complete
basis set, we have to shift perspectives how to view the NRG
algorithm.

There are two possible ways to interpret the iterative
NRG solution of the $N$-site chain. In the traditional
picture one starts from a core cluster that consists of
the impurity degrees of freedom and the $N = 0$ site,
and enlarges the chain by one site at each NRG step.
Alternatively, one can view the NRG procedure as
starting from the full chain of length $N$, but with
the hopping matrix elements set to zero along the chain.
At each successive step another hopping matrix element is
switched on, until the full Hamiltonian ${\cal H}_N$ is
recovered. In this latter picture, to be adopted below,
the entire sequence of Hamiltonians ${\cal H}_m$ with
$m \leq N$ act on the Fock space of the $N$-site chain.

Accordingly, each NRG eigen-energy of ${\cal H}_m$ has
an extra degeneracy of $d^{(N-m)}$, where $d$ is the
number of distinct states at each site along the chain.
The extra degeneracy stems from the $N-m$ ``environment''
sites at the end of the chain, depicted in
\begin{figure}[htbp]
\centering
\includegraphics[width=80mm,clip]{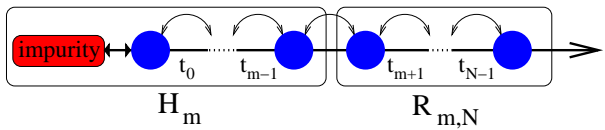}
\caption{(color online) The full Wilson chain of length $N$ is divided into
         a sub-chain of length $m$ and the ``environment''
         $R_{m,N}$. The Hamiltonian ${\cal H}_m$ can be
         viewed either as acting only on the sub-chain
         of length $m$, or as acting on the full chain of
         length $N$, but with the hopping matrix elements
         $t_m,\cdots,t_{N-1}$ all set to zero. The former
         picture is the traditional one. In this paper we
         adopt the latter point of view.} 
\label{fig:1}
\end{figure}
Fig.~\ref{fig:1}.

The set of  eigenstates of  ${\cal H}_m$,  conventionally denoted as
$\{ \ket{r} \}$, can be 
formally constructed from the complete basis set $\{
\ket{\alpha_{imp},\alpha_0,\cdots, \alpha_{N} } \}$ of the NRG chain
of length $N$ where the $\alpha_i$
label the configurations on each chain link $i$. Since ${\cal H}_m$
does not act on the chain links $m+1,\cdots,N$, $\ket{r}$ can be
written as $\ket{r,e;m}$  where the ``environment''
variable $e = \{\alpha_{m+1}, \cdots,\alpha_N\}$ encodes the
$N - m$ site labels $\alpha_{m+1}, \cdots, \alpha_N$.
 The index $m$ is used in this notation to record where the chain
is partitioned into a ``subsystem'' and an ``environment''
(see Fig.~\ref{fig:1}).

Consider now the first iteration $m_{\rm min}$ at which states are
discarded. In order to keep track of the complete basis
set of the $N$-site chain, we formally divide the
eigenstates $\ket{r,e;m}$ into two distinct subsets: the
discarded high-energy states $\{ \ket{l,e;m}_{dis} \}$ and
the retained low-energy states $\{ \ket{k,e;m}_{kp} \}$.
In the course of the NRG, only the latter states are used
to construct the next  Hamilton matrix $\H_{m+1}$ within the reduced
subspace $\{ \ket{k,\alpha_{m+1},e';m} \}$. Note, however,
that the sum of both subsets still constitutes a complete
basis set for the $N$-site chain.
Repeating this procedure at each subsequent iteration, we
recursively divide the retained subset into a discarded
part and a retained part. Then, the collection of {\em all
discarded} eigenstates $\ket{l,e;m}_{dis}$ together with
the eigenstates of the final NRG iteration $N$ combine
to form a complete basis set for the entire Fock space
${\cal F}_{N}$. Regarding all eigenstates of the final
NRG iteration as discarded, one can formally write the
Fock space of the $N$-site chain in the form
${\cal F}_{N} = {\rm span} \{|l,e;m \rangle_{dis}\}$.
Since {\em all states} are retained in the course
of this construct, the following completeness relation
obviously holds:
\begin{eqnarray}
\sum_{m = m_{\rm min}}^N \sum_{l,e}
          \ket{l,e;m}_{dis}\ _{dis}\bra{l,e;m} &=& 1 \; .
\label{equ:complete-basis}
\end{eqnarray}
Here the summation over $m$ starts from the first iteration
$m_{\rm min}$ at which a basis-set reduction is imposed.
All traces  below will be carried out with respect to
this basis set. Hence, {\em the evaluation of the spectral functions
  will not involve any truncation error}. Note also that we
made no reference to a particular Hamiltonian ${\cal H}$ 
in constructing the basis set $\{ |l,e;m \rangle_{dis} \}$.

Another useful identity to be used below pertains to the
subspace retained at iteration $m$, $\{ \ket{k,e;m}_{kp}\}$. To
this end, we note that the sum in Eq.~(\ref{equ:complete-basis})
can always be divided into two complementary parts $1_m^-$
and $1_m^+$: 
\begin{eqnarray}
1_m^- &=& \sum_{m'=m_{min}}^{m} \sum_{l',e'}
               \ket{l',e';m'}_{dis}\ _{dis}\bra{l',e';m'} \; ,
\label{eqn:partition-fock-space-i}
%%\label{eqn:1_m^-}
\\
1_m^+ &=& \sum_{m'= m+1}^N \sum_{l',e'}
               \ket{l',e';m'}_{dis}\ _{dis}\bra{l',e';m'} \; .
\nonumber \\
& =& \sum_{k,e} \ket{k,e;m}_{kp} \, _{kp}\bra{k,e;m} \; .
\label{eqn:partition-fock-space-ii}
\label{eqn:partition-fock-space}
%%\label{eqn:1_m^+initial}
\end{eqnarray}
with the  completeness relation
\begin{eqnarray}
  1 &=& 1_m^- + 1_m^+ \; .
\label{eqn:completness}
\end{eqnarray}
Note that for $m = N$ only $1_m^-$ exists.

\subsection{Definition of the Green's function} 

The textbook definition of the retarded Green's function 
is given by
\begin{equation}
\label{eqn:6}
  G_{A,B}(t) = -i\Theta(t)\Tr{\hat{\rho} \left[A(t) B\right]_{-s}}
\end{equation}
where $[A,B]_s=AB-sBA$ with $s=1$ for bosonic operators $A,B$ and $s=-1$ for
fermionic operators. In our investigation, we restrict ourselves to
local operators $A$ and $B$.\cite{AndersSchiller2005} 

As already mentioned, the fundamental philosophy of the NRG is that a
chain of length $N$ corresponds to a temperature scale $T\sim D_N$, hence
$\beta E_l^m\gg1$ for all $m<N$, where $E_l^m$ denotes the eigen energies
of ${\cal H}_m$.
In the NRG, the
thermodynamic density operator $\hat{\rho}$ is therefore represented only by the
states of the last iteration $N$ 
\begin{eqnarray}
\label{eqn:rho-nrg}
  \hat{\rho} &=& \frac{1}{Z} e^{-\beta H} \approx \frac{1}{Z_N} \sum_l
  \rho_l \ket{l;N} \bra{l;N} \\ 
\rho_l &=& \frac{e^{-\beta E_l^N}}{Z_N}
\end{eqnarray}
and $Z_N=\sum_l e^{-\beta E_l^N}$.

If we were able to solve the NRG chain of length $N$ without any
truncation, the Green's function $G_{A,B} (z)$ would be
given by the textbook Lehmann representation
\begin{eqnarray}
  G_{A,B} (z) &=& \int_0^\infty dt e^{izt}  G_{A,B} (t) 
\non
&=&
\frac{1}{Z} \sum_{l,l'} A_{ll'} B_{l'l}
\frac{e^{-\beta E_l} - s \,e^{-\beta E_{l'}}}{z  +E_l-E_{l'}}
\label{eqn:lehmann-representation}
\end{eqnarray}

\subsection{Derivation of the NRG Green's function}
\label{sec:II.C}

Let us start from the retarded Green's function looking only at the first
term from the commutator
and insert two completeness  relations (\ref{eqn:completness}). We then obtain
\begin{widetext}
\begin{eqnarray}
\label{eqn:9}
\Tr{\hat{\rho} e^{iHt} A e^{-iHt} B }
 &=& 
 \sum_{l,e,m}
 \sum_{l',e',m'}
 \Tr{\hat \rho e^{iHt}\ket{l,e;m}\bra{l,e;m} A e^{-iHt} 
\ket{l',e';m'}\bra{l',e';m'}
B }
\nonumber\\
&=&
\label{eqn:gf-three-terms}
\sum_m \sum_{l,e}
 \sum_{l',e'} \, _{dis}
\bra{l,e;m} A e^{-iHt} \ket{l',e';m}_{dis} \,_{dis}\bra{l',e';m}
B  \hat{\rho} e^{iHt}\ket{l,e;m}_{dis}
\non
&& +
\sum_m \sum_{l,e}
 \sum_{k,e'} \,  _{dis}
\bra{l,e;m} A e^{-iHt}  \ket{k,e';m}
_{kp} \,  _{kp}
\bra{k,e';m}
B \hat \rho e^{iHt}\ket{l,e;m}_{dis}
\non
&&
+
\sum_{m} 
 \sum_{l,e'}
\sum_{k,e} \, _{dis}
\bra{l,e';m} B e^{iHt}\hat \rho\ket{k,e;m}
_{kp} \,  _{kp}
\bra{k,e;m} A e^{-iHt }  \ket{l,e';m}_{dis}
\;.
%\nonumber
\end{eqnarray}
\end{widetext}
For the second term from the commutator in (\ref{eqn:6}) one
obtains a similar expression.
Here and in the following we use the convention that an index
$l$ labels a discarded state, while an index $k$ represents a state
kept
at a certain chain length $m$.
The first term in (\ref{eqn:gf-three-terms}) arises from $m'=m$, the second from $m'>m$ and the third
from $m'<m$, where we made use of Eq.~(\ref{eqn:partition-fock-space}).
This trick\cite{AndersSchiller2005} allows to express the
Green's function as sum over equal shell contributions only. Note
that this is an exact formula: {\em no approximations} have been made
so far!

Since the state $ \ket{s,e;m}$ is an eigenstate of $H_m$, i.e.\ $H_m
\ket{s,e;m} = E_s^m  \ket{s,e;m}$, we will now use the approximation
$H\ket{s,e;m} \approx E_s^m  \ket{s,e;m}$. This approximation,
justified due to the energy hierarchy implied by the logarithmic
discretization,\footnote{In contrast to the standard NRG notations we 
consider $\H_N$ as dimensionfull Hamiltonian in order to kept the
notation as simple as possible.} is
completely in the spirit of the NRG and in fact used in the calculation
of thermodynamics properties.\cite{Wilson75} 
Note, that this will be the only approximation made, which is of
energetic nature and unrelated to any truncation error. We thus
evaluate $\exp(iHt) \ket{s,e;m} 
\approx \exp(iE_s^mt) \ket{s,e;m}$ and Laplace transform all
contributions. The first term from (\ref{eqn:gf-three-terms}) and the
corresponding expression for the second term from the commutator 
contains only discarded states  and reduces, therefore, to the last iteration $m=N$
due to the representation of the density operator in NRG by Eq.\
(\ref{eqn:rho-nrg}). It yields as contribution to $G_{A,B}(z)$
\begin{eqnarray}
\label{eqn:green-i}
  G^{i}_{A,B}(z)
&=&
\frac{1}{Z}\sum_{l,l'}
\bra{l;N} A  \ket{l';N}\bra{l';N}
B \ket{l;N}
\non
&& \times
\frac{e^{-\beta E^N_l}-s \,e^{-\beta E^N_{l'}}}{z +E^N_{l}-E^N_{l'}}
\end{eqnarray}
In the next two terms the summation over all energy shells has to be
evaluated, because the states $\ket{k,e;m}_{kp}$ are not orthogonal to
$\ket{l;N}$. We obtain, using $\hat\rho\ket{l,e;m}_{dis}=0$ for $m<N$,\cite{AndersSchiller2006}
\begin{widetext}
\begin{eqnarray}
  G^{ii}_{A,B}(z)
&=&
\sum_{m=m_{min}}^{N-1}
\sum_{l,e}
 \sum_{k,e'}
\bra{l,e;m} A \hat{\rho}   \ket{k,e';m}\bra{k,e';m}
B \ket{l,e;m}
%%\non
%%&& \times
\frac{-s}{z +E^m_{l}-E^m_{k}}
\end{eqnarray}
and
\begin{eqnarray}
  G^{iii}_{A,B}(z)
&=&
\sum_{m=m_{min}}^{N-1}
 \sum_{l,e'}
\sum_{k,e}
\bra{l,e';m} B \hat{\rho} \ket{k,e;m}
\bra{k,e;m} A  \ket{l,e';m}
%%\non
%%&& \times
 \frac{1}{z +E^m_{k}-E^m_{l}}
\end{eqnarray}
\end{widetext}
Since in the last iteration there are no kept states, all are
considered discarded, $m=N$ does not contribute to $G^{ii}$ and
$G^{iii}$. Now we insert a completeness  relation $(1^+_m +1_m^-)$
between $A$ and the density operator in $G^{ii}_{A,B}(z)$ and $B$ and the density 
operator in $G^{iii}_{A,B}(z)$ and make use of $1_{m}^- \hat{\rho} =
0$ for all $m<N$.\cite{AndersSchiller2006} Because  $A$ and $B$ are local
operators,\footnote{Local operators we call operators which act only on chain
  links up to $m<m_{min}$. See a detailed discussion in Refs.\
  \onlinecite{AndersSchiller2005,AndersSchiller2006}.
}
the matrix elements \begin{equation}
  \bra{k,e;m} A \ket{l,e';m} =
s^{n_{e'}}
\delta_{e,e'} 
A_{k,l}(m)
\end{equation}
are diagonal and independent of the environment variables $e$. $n_{e'}$
denotes the number of Fermions in the environment times the total
number of Fermions created by $\hat A$. Since the matrix
elements of $A$ and $B$ are evaluated 
simultaneously, the total
% fermionic 
phase factor is given by
$[ s^{n_{e'}}]^2 =1$ for operator $A$ and $B^\dagger$ imposing the
same change of the total particle number. 
Therefore, the trace over the environment only 
acts on the density operator $\hat{\rho}$ and the final result
\begin{eqnarray}
  G^{ii}_{A,B}(z)
&=&
\sum_{m=m_{min}}^{N-1}
 \sum_{l}
 \sum_{k,k'} A_{l,k'}(m) \rho^\textrm{red}_{k',k}(m)B_{k,l}(m)
\non
&& \times 
\frac{-s}{z +E_{l}-E_{k}}
\label{eqn:green-ii}
\end{eqnarray}
and
\begin{eqnarray}
  G^{iii}_{A,B}(z)
&=&
\sum_{m=m_{min}}^{N-1}
 \sum_{l}
\sum_{k,k'} B_{l,k'}(m)\rho^\textrm{red}_{k',k}(m) A_{k,l}(m)
\non
&&
\times
 \frac{1}{z +E_{k}-E_{l}}
\label{eqn:green-iii}
\\
G_{A,B}(z) &=& G_{A,B}^{i}(z)+G_{A,B}^{ii}(z)+G_{A,B}^{iii}(z)
\label{eqn:green-tot}
\end{eqnarray}
is formulated in terms of the reduced density matrix\cite{Feynman72,White92}
\begin{equation}
\rho^{\rm red}_{k,k'}(m) = \sum_{e}
          \langle k,e;m|\hat{\rho} |k',e;m \rangle \; ,
\label{eqn:reduced-dm-def}
\end{equation}
first introduced to the calculations of NRG spectral functions by
Hofstetter.\cite{Hofstetter2000}  The part
$G^{ii}(z)$ describes negative frequency excitations, while
$G^{iii}(z)$ accounts for positive frequency excitations for all
$m<N$ because $E_l-E_k>0$ by construction.  $G^{i}(z)$ sums all
excitations of the last iteration $N$ and has the form of the usual
Lehmann representation Eq.\ (\ref{eqn:lehmann-representation}).

A few words are in place to pinpoint the difference to the Hofstetter
approach\cite{Hofstetter2000} to calculate spectral functions with NRG. Although our
reduced density matrix $\rho^{\rm red}$ is identical to the one given
in Eq.~(7) in Ref.~\onlinecite{Hofstetter2000}, our rigorous derivation
differs in the summation of excitations contributing to the Green's
function. Eqs.\ (\ref{eqn:green-ii}) 
and (\ref{eqn:green-iii}) state clearly that one {\em must} only
include excitations between a {\em discarded} and {\em kept} state
while in the  Hofstetter approach\cite{Hofstetter2000} the summation
index $l$ runs over all states present at iteration $m$. This leads
to an overcounting of contributions, by the way inherent to all previous approaches to NRG
spectral
functions, see for details Refs.\
\onlinecite{CostiHewsonZlatic94,Costi97,BullaHewsonPruschke98,Hofstetter2000,BullaCostiVollhardt01}
and  references therein. The origin
of the restriction of summation in Eqs.\
(\ref{eqn:green-ii}) and (\ref{eqn:green-ii}) is quite  
obvious: All kept states in iteration $m$ span the Fock-space of
Hamiltonian $H_{m+1}$ and, therefore, will contribute to the excitations at a
latter iteration $m'>m$. They {\em must not} be included at iteration $m$.

In addition, the phenomenological patching
algorithms,\cite{BullaCostiVollhardt01} where spectral information from
different energy shells are ``merged'' become obsolete in our
approach. Equations (\ref{eqn:green-i}-\ref{eqn:green-iii}) state exactly which
excitations contribute at which Wilson shell. Obeying this summation
restriction ensures the fulfillment of the spectral sum rules
independent of the number $N_S$ of NRG states kept at each
iteration. With a little algebra and the completeness relations
(\ref{eqn:completness}), we show in appendix \ref{sec:a1} that the
spectral sum rule 
\begin{eqnarray}
C &=& \oint \frac{dz}{2\pi i} G_{A,B}(z) 
\non
&=&  \sum_{\alpha=i,ii,iii} \oint \frac{dz}{2\pi i} G^{\alpha}_{A,B}(z) =
  \Tr{\rho [A,B]_{s}}
\label{eqn:contour})
\end{eqnarray}
is always fulfilled {\em exactly}.

Therefore, the  spectral function will be correctly normalized independent of the
number of states $N_S$ kept after each NRG iteration. Note that the NRG
truncation only influences the 
partitioning of the states, but never the completeness of the basis.
Hence, the spectral functions become more robust to truncation
errors,  as will be shown in section \ref{sec:results} by using an
extremely low number of kept NRG states. The spectral sum rule is a
consequence of an operator identity and independent of approximations
made in the dynamics as long as {\em no states} are discarded in the
calculation.

\subsection{Occupation number}
\label{sec:occ}

Let us specialize to the local spin-dependent fermionic spectral
function, i.e.\ $A=f_\sigma$ and $B=f^\dagger_\sigma$. The local
occupation $\expect{f_\sigma^\dagger f_\sigma}$ can be expressed with
the spectral integral as
\begin{eqnarray}
  \expect{f_\sigma^\dagger f_\sigma} &=&
 \int_{-\infty}^\infty
  \frac{d\w}{\pi} f(\w) \Im mG_{f_\sigma,f_\sigma^\dagger}(\w -i\delta)
\non
&=&
 \oint   \frac{d z}{2\pi i} f(z) G_{f_\sigma,f_\sigma^\dagger}(z)  
\label{equ:occupation-sum-rule}
\end{eqnarray}
where $f(\w)$ is the Fermi function. By substituting Eqs.\
(\ref{eqn:green-i}-\ref{eqn:green-iii}) for
$G_{f_\sigma,f_\sigma^\dagger}(z)$ and evaluating the contour
integration, we show in appendix \ref{app:occ-sum-rule}  explicitly
that our expressions approximately -- at $T=0$ even exactly -- reproduce 
the expectation value
$\expect{f_\sigma^\dagger f_\sigma}$ calculated directly with the NRG. 
In contrast to the spectral sum rule, however,  which
is {\em exact} and, therefore, reproduced in our numerics with
machine accuracy, the accuracy of the occupation numbers calculated
from the spectra depends on the validity of the assumption of
vanishing  Boltzmann factors. Therefore, we expect it to show a
certain error at finite temperatures. Nevertheless, we find that the
deviation between the NRG 
values and  the ones obtained by the numeric evaluation\footnote{We
  can either directly sum the weights of the discrete raw NRG spectrum or
  perform a numerical integration after having broadened it. The former
  usually yields an order of magnitude better accuracy than the latter.} of
(\ref{equ:occupation-sum-rule}) remains less 
than $10^{-4}$. This implies also that the NRG value for quantities
like the local magnetization $m= \sum_\sigma \sigma
\expect{f_\sigma^\dagger f_\sigma}/2$   is {\em accurately} reproduced by our formulation of
the Green's function. This is a significant
improvement over the Hofstetter approach, where deviations on the 
percent level between
$m^{NRG}$ and $m^{GF}$ have been reported.\cite{Hofstetter2000}

\section{Results}
\label{sec:results}

\subsection{Single impurity Anderson model}

The general scheme for the calculation of spectral functions presented in
section \ref{sec:theory} did not make any reference to a certain model or
bath statistics. Therefore, our algorithm is suitable for any quantum
impurity system. In order to demonstrate the virtue of the new
approach, we will present in this section as an important example calculations
for the spin-dependent single-particle spectral functions of the
single impurity Anderson model (SIAM) with and without an external
magnetic field at $T=0$ and finite $T$.

The
Hamiltonian of the SIAM\cite{Anderson61,KrishWilWilson80a,KrishWilWilson80b}
\begin{eqnarray}
  \H &=& \sum_{k\sigma} \e_{k\sigma} c^\dagger_{k\sigma}c_{k\sigma}
+ \sum_{\sigma} \left(\e_f -\sigma H\right) f^\dagger_\sigma f_\sigma
\non
& & + \frac{U}{2}\sum_\sigma n^f_{\sigma} n^f_{-\sigma}
+ V\sum_{k\sigma}  \left( c^\dagger_{k\sigma} f_\sigma
+f^\dagger_{\sigma} c_{k\sigma}\right)
\end{eqnarray}
consists of a single local state, which we will denote with $f$, with energy
$\e_f$ and Coulomb repulsion $U$,
coupled to a bath of conduction electrons with creation operators
$c^\dagger_{k\sigma}$ and energies $\e_{k\sigma}$. The local level is
subject to a Zeemann splitting in an external magnetic field $B$.
Note that we denote with $H=g\mu_{\rm B}B/2$ the Zeeman energy and the
total splitting $\left|\epsilon_\uparrow-\epsilon_\downarrow\right|=2H$.
We employ Wilson' s  NRG\cite{KrishWilWilson80a,KrishWilWilson80b} to
generate the eigen-energies, the matrix elements and basis set needed
for the Green's function.

So far, all analytical calculations were performed using the discrete
NRG spectrum. To obtain a continuous spectral function  from the
set of discrete $\delta$-functions occurring in $G_{A,B}(z)$, we
have to introduce a coarse-graining. Due to the exponentially
decreasing energy scales, a broadening on a logarithmic mesh by a Gaussian
\begin{equation}
  \delta(\w -\w_n) \to \frac{e^{-b^2/4}}{b\w_n\sqrt{\pi}} \exp\left\{-\left(\frac{\ln(\w/\w_n)}{b}\right)^2\right\}
\label{eqn:broadening}
\end{equation}
is typically used.\cite{SakaiShimizuKasuya1989,CostiHewsonZlatic94} Since this
broadening function is properly normalized to one, the spectral weight 
is conserved and no principle inaccuracies are introduced by this procedure.

\subsection{Spectral functions for $T=0$}

In the following, we specialize to $A=f_\sigma$ and $B=f^\dagger_\sigma$.
For comparison we calculated the raw NRG spectral function in three different ways:
(i) by the conventional way as described in Refs.\
\onlinecite{CostiHewsonZlatic94,BullaCostiVollhardt01} labelled as
(CON) in the following, (ii)  by the
Hofstetter approach\cite{Hofstetter2000} (HA) 
and by our method defined by Eqs.\
(\ref{eqn:green-i})-(\ref{eqn:green-tot}) labelled {\em complete Fock
  space} approach (CFS).
As long as not stated otherwise, all energies are measured in units of
the half band width $D$, and for simplicity only a symmetric conduction band
is considered  with a constant density of states $\rho_0=1/(2D)\Theta(D-|\omega|)$.\cite{Wilson75}

\begin{figure}[htb]
  \centering
  \includegraphics[width=80mm,clip]{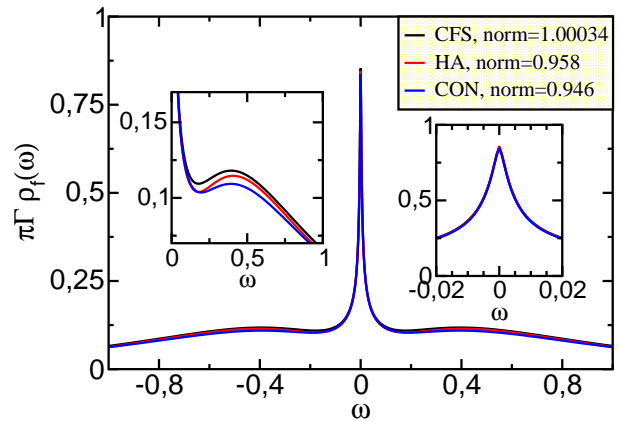}
  \caption{(color online) Comparison of the spectral function for the three different
    methods for the symmetric case $\e_f/D=-U/D2=0.5$. The inset on
    the left side emphasizes the broad charge excitation at $\e_f$,
    the  inset on the right side zooms in at the Abrikosov-Suhl
    resonance (ASR). NRG parameters:  $\Gamma/D =\pi V^2\rho_0/D=0.1,
    \Lambda=2.5, N_{S}=500, b=0.8, T= 0$} 
  \label{fig:2}
\end{figure}

In Fig \ref{fig:2} we show a comparison of spectral functions
calculated by the three different methods using identical NRG input
data and broadening parameters. All three methods agree on the shape
and the height of the 
many-body resonance (ASR) at the Fermi energy and all methods  do
not reach the unitary limit of $\pi\Gamma \rho(0)=1$ as predicted by the
density of states rule.\cite{AndersGreweLorek91} The spectral
functions, however, differ in their high energy features. The
conventional spectral function underestimates the charge excitations and only
reaches a total spectral weight of $C=0.946$. The Hofstetter
approach\cite{Hofstetter2000} yields a somewhat improved value
of $C=0.956$ because the high energy part of its spectrum, which carries
most of the spectral weight, lies between our and the conventional curve. The
spectral function based upon our new  approach, however, reaches the
spectral weight $C=1.00034$. It deviates from the exact value of $C=1$
only by the error introduced by the numerical $\w$ integration. To verify this, we
add all spectral weights from Eqs.\ (\ref{eqn:green-i})-(\ref{eqn:green-iii})
directly and obtain $|C-1|=10^{-12}\ldots10^{-15}$, i.e.\ the norm is
$C=1$ within machine accuracy.\footnote{This in fact holds for
  all spectra calculated, i.e.\ with or without magnetic field, at
  $T=0$ and finite temperature!}
For this symmetric case, its obvious that our approach also yields the
{\em exact} occupation number of $n_\sigma^f = \expect{\hat
  n^f_\sigma} = 0.5$ for each spin direction while the other two
approaches have a $5.7\%$ (CON) or $4.6\%$ (HA) error. Note, however,
that the latter two still give the correct value $C/2$ with respect to
their respective norm.

At this point a comment on the violation of the density of
states rule  $\pi\Gamma \rho(0)=1$ for $T\to 0$
seems in order. Since the NRG itself
does only provide the weight of $\delta$-peaks,  the height of the
coarse-grained spectrum will depend on the density of these
peaks, i.e.\ the number of states available in the calculation, and
the choosen broadening function. This {\em does not} imply the
violation of Friedel sum rule\cite{AndersGreweLorek91}, too, which
relates the {\em scattering phases} to the number of displaced 
electrons in each spin channel.  Since the NRG fixed point spectrum contains the correct
scattering phases, it is accurately fulfilled. However, the relation
between the phase shift and the local spectral function is in general
not very accurately reproduced due to the aforementioned reasons.
Nevertheless,  the density of states sum rule can be
satisfied with an error of less than $3\%$ using small values of
$\Lambda$, more NRG states and a smaller
broadening\cite{CostiHewsonZlatic94} 
as used in  Fig \ref{fig:2}.

To our knowledge, only the indirect calculation of
the physical Green's function by expressing its self-energy by the ratio
of two correlation functions\cite{BullaHewsonPruschke98} yields
values which are accurate enough to reproduce
the correct height of $\rho(0,T=0)$ nearly independent of the
broadening parameters. In particular, this is of importance to prove
the pinning of the spectral function for the two-channel Anderson
model at half the unitary limit.\cite{AndersTCSIAM2005}

 \begin{figure}[htb]
   \centering
   \includegraphics[width=80mm,clip]{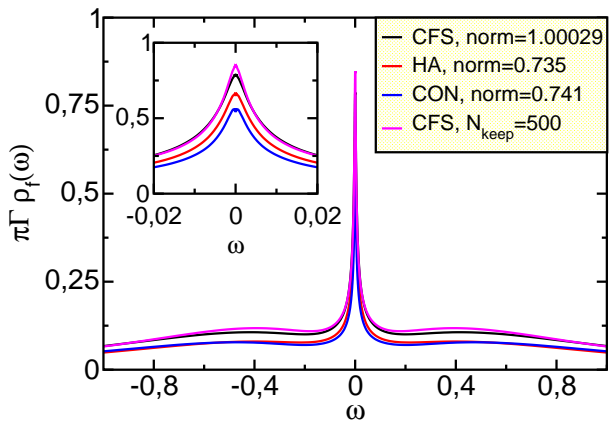}
   \caption{(color online) Comparison of the spectral function for the three different
     methods for the same model and NRG parameters as in
     Fig.~\ref{fig:2} but with only $N_S=50$ states kept in each NRG
     iteration. The inset zooms in onto the  ASR.
   \label{fig:3}
   }
 \end{figure}
We have put our method to an extreme test by reducing the number of
states $N_S$ kept after each NRG step to $N_S=50$ for the same
$\Lambda=2.5$ as before. This is definitely  stretching the limit of
the conventional NRG beyond its accuracy. The results for the spectral
functions are plotted in Fig.~\ref{fig:3}. Only data for
frequency $|\w|>10^{-7}$ are included, which, however, is much smaller
than the Kondo temperature $T_K$ as estimated from the width of the
ASR. The inaccuracy of the matrix elements for extremely small
frequencies $|\w|<10^{-7}$  yields erratic results in this
interval which, therefore, are omitted here. While the
conventional and the HA  methods significantly loose 
spectral weight, the spectral weight of CFS spectral function remains
at the exact value $C=1$. Since the discretization of the energy mesh
was the same in all figures, the deviation from the sum rule of
$3\times 10^{-4}$ again stems only from the numerical integration. For
comparison we also added the CFS curve from Fig.~\ref{fig:2}, calculated
with $N_S =500$ NRG states. Both CFS curves agree extremely well. In
particular, the weight under the charge peak is only slightly redistributed
and the shape of
the ASR remains unaltered. Only at very low frequency, the inaccuracy
of the underlying NRG input data takes its toll: the ASR peak height is
further reduced. The other two methods, however, exhibit strong
dependence on $N_S$ in all energy regimes. We must report an error in the  spectral weight
and the occupancy $n_\sigma^f$ of $\approx 35\%$ for both the
conventional and the HA method. In all cases the same broadening
parameter $b=0.8$ was used.

\begin{figure}[htb]
  \centering
  \includegraphics[width=80mm,clip]{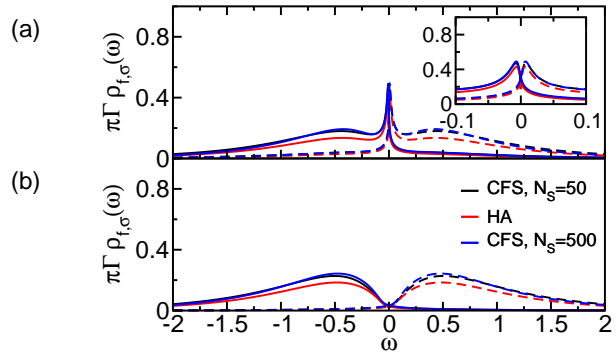}
  \caption{(color online) Comparison of the spectral function obtained from CFS and
    HA approaches in a finite magnetic field (a) $H=0.005$ and (b)
    $H=0.1$. The full curves represent spin up, the dashed ones spin
    down. The inset in (a) is a zoom into the region around the Fermi level
    showing the splitting and suppression of the Kondo resonance. Model and NRG parameters as in Fig.~\ref{fig:3}. The norm
    and occupancies obtained are compared in Tab.~\ref{tab:symm_with_b}
    }
  \label{fig:4}
\end{figure}
While the conventional method yields quite reliable spectral
information for small $\Lambda$ and large number of states in the
absence of a magnetic field, it fails for finite 
magnetic field, where the fixed point for $T\to 0$ yields a complete
redistribution of spectral weight on all energy scales compared to the
spectral function\cite{Hofstetter2000} at $H=0$. Therefore, it is
omitted in  Fig.~\ref{fig:4}, where we compare spectral functions for two
magnetic fields, $H=0.005\approx T_K$ (Fig.~\ref{fig:4}a) and $H=0.1\gg T_K$
(Fig.~\ref{fig:4}b); the results for the majority spin are plotted as solid
lines, the ones for the minority spin as dashed lines. For $H\approx T_K$ the
ASR is split and already reduced to half its original height, while for
$H\gg T_K$ it has vanished completely, as expected.  Again, we
display the data for $N_S=50$ to highlight the differences in the 
methods.  While our method and the HA approach exhibit the same
overall features, i.e.\ splitting and reduction respectively complete lack of
the ASR and a major redistribution of
spectral weight, the charge excitation peak contains much more
spectral weight in the CFS as the HA curve. As before, the CFS spectral function
\begin{table}[htb]
\setlength{\tabcolsep}{1mm}
\centering
\begin{tabular}{llrrrrr}
 & \multicolumn{1}{c}{Method} 
 & \multicolumn{1}{c}{Norm} 
 & \multicolumn{2}{c}{$\langle n_\sigma\rangle_{GF}$} 
 & \multicolumn{2}{c}{$\langle n_\sigma\rangle_{NRG}$}\\
 &
 &
 & \multicolumn{1}{c}{$\sigma=\uparrow$} 
 & \multicolumn{1}{c}{$\sigma=\downarrow$} 
 & \multicolumn{1}{c}{$\sigma=\uparrow$} 
 & \multicolumn{1}{c}{$\sigma=\downarrow$} \\
\hline 
(a) & CFS(500) & 1 & 0.807 & 0.193 & 0.807 & 0.193\\
%\cline{2-7}
 $H=0.005$ & CFS(50) & 1 & 0.806 & 0.194 & 0.806 & 0.194\\
%\cline{2-5}
 & HA & 0.747 & 0.600 & 0.147 & 0.806 & 0.194\\
\hline
%\hline
(b) & CFS(500) & 1 & 0.953 & 0.047 & 0.953&0.047\\
%\cline{2-5}
$H=0.1$ & CFS(50) & 1 & 0.953 & 0.047 & 0.953 & 0.047\\
%\cline{2-5}
 & HA & 0.793 & 0.755 & 0.038 & 0.953 & 0.047\\
\hline
\end{tabular}
\caption{Norm and occupancies for the calculations with finite magnetic field
in Fig.~\ref{fig:4}. The last columns contain the occupancies obtained from
the thermodynamic expectation values.\label{tab:symm_with_b}}
\end{table}
fulfills the spectral sum rule with the accuracy of $10^{-4}$ while we
note a $20\%$ error in the HA curve for these extreme set of
parameters. The CFS results for $N_S=500$ given by the blue curves in Fig.~\ref{fig:4}
by and large again lie on top of the CFS curves for $N_S=50$, once more
emphasizing that within the CFS neither  the accuracy of the occupancy
nor the one of the spectral functions does critically depend on the
number of states kept.

This feature becomes even more apparent from the actual numbers for the norm
and the spin-dependent occupation numbers listed in
Tab.~\ref{tab:symm_with_b}. The last column shows for comparison the
occupancies for the two spin directions as calculated directly from
the NRG level flow.\cite{Wilson75} Note that the difference in the
occupancies between $N_S=500$ and $N_S=50$ in the CFS appear in the
thermodynamic occupation numbers, too. 
On the other hand, there exist significant differences in the magnetization
$m=(n_\uparrow -n_\downarrow)/2$ obtained from the HA method, which yields a
$30\%$ error compared to the reference NRG magnetization and our CFS method.

We like to emphasize that  the rather large error of the HA
method results from the unusual low number of NRG state $N_S$ kept
after each NRG  iteration. Increasing the number of states and
reducing $\Lambda$ one can increase the accuracy of the HA method
considerably, but typically is always left with about $4\%$ error according to
Tab.~I in Ref.~\onlinecite{Hofstetter2000}. Our motivation for choosing
$N_S=50$ was to put our initial claim to an extreme test: The spectral sum rule
and the occupation numbers are reproduced with high accuracy {\em independent} of
the number $N_S$ of states kept after each iteration. 
{\em Our equations  (\ref{eqn:green-i})-(\ref{eqn:green-iii}) do not
  contain any truncation errors since we use a complete basis set.}
The errors in the CFS spectral functions are of purely energetic nature and
will indeed be found in the thermodynamic quantities, too. As a
consequence, the spectral functions appear to be largely insensitive
to the number of NRG state kept.

 \begin{figure}[htb]
   \centering
   \includegraphics[width=80mm,clip]{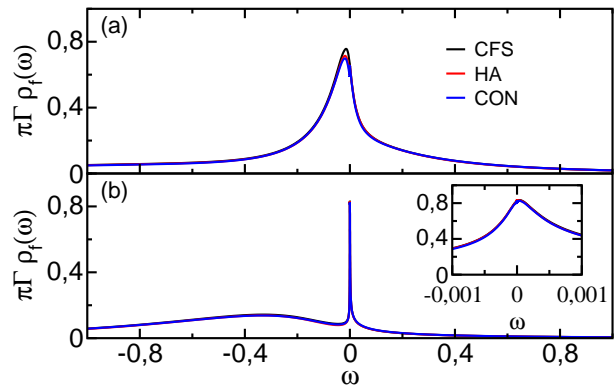}

   \caption{(color online) Comparison of the spectral function for the three different
     methods for the asymmetric case (a) $U/D=1$ and $\e_f/D= -0.9$,
     and (b) $U/D=1000$ and $\e_f/D=-0.4$. NRG
     parameters: $\Gamma/D =\pi V^2\rho_0/D=0.1, \Lambda=2.5,
     N_{S}=500$. The results for norm and occupancies obtained are
   collected in Tab.~\ref{tab:asymmetric}.}
   \label{fig:5}
 \end{figure}
We made the same observation as discussed above in the asymmetric case
shown in Fig.~\ref{fig:5}(a). The overall shape of the spectral functions of
all three methods agree pretty well. Again, we note differences in the
deviations from the sum rules, see Tab.~\ref{tab:asymmetric}. While our new CFS method obeys the
\begin{table}[htb]
\setlength{\tabcolsep}{2mm}
\centering
\begin{tabular}{llrrr}
 & \multicolumn{1}{c}{Method} & \multicolumn{1}{c}{Norm} &
 \multicolumn{1}{c}{$\langle n\rangle_{GF}$} & \multicolumn{1}{c}{$\langle n\rangle_{NRG}$}\\
\hline 
(a) & CFS & 1 & 0.666 & 0.666\\
%\cline{2-4}
$U=1$ & HA & 0.974 & 0.645 & 0.666\\
%\cline{2-4}
 & CON & 0.975 & 0.645 & 0.666\\
\hline
%\hline
(b) & CFS & 0.543 & 0.915 & 0.915\\ 
%\cline{2-4}
$U=\infty$ &  HA & 0.523 & 0.878 & 0.915\\
%\cline{2-4}
 & CON & 0.534 & 0.893 & 0.915\\
\hline
\end{tabular}
\caption{Norm and occupancies obtained with the different methods to
calculate spectra for the asymmetric SIAM (see Fig.~\ref{fig:5} for
the parameters used). The last column shows the occupancy as obtained
from the thermodynamic expectation value.\label{tab:asymmetric}}
\end{table}
spectral weight sum rules {\em and} the occupation sum rule very
accurately, the comparison of the actual numbers for norm and occupancy 
in Tab.~\ref{tab:asymmetric}(a) shows that the conventional and the HA method yields $C=0.974$ and the
Green's function occupancy of $n_\sigma^{GF}=0.645$ deviates from the
NRG occupancy $n^{NRG}_\sigma = 0.666$ by $3\%$. 

In Fig.~\ref{fig:5}(b) the spectral functions for $U/D=1000$ are
plotted. This corresponds to the limit $U\to \infty$ of the SIAM,
i.e.\ the charge excitation between the singly and the doubly 
occupied state is effectively shifted to infinity. Therefore, the total spectral
weight is reduced to $C=1-\expect{\hat n_f}/2$, if we neglect the
$\delta$-like peak at $\w\approx U$. Nevertheless, adding all discrete
NRG spectral weights yields $|C-1|\approx 10^{-12}-10^{-15}$ even in
this case. Again, the values in 
Tab.~\ref{tab:asymmetric}(b) show that the CFS
spectral function agrees excellently with the exact norm but show a
slight error of $0.03\%$ in the occupancy, which we again attribute to
the numerical integration, while the conventional and
the HA methods have errors at least two orders of magnitude larger of about $4\%$.

\subsection{Finite temperature spectral functions} 
\label{sec:finite_T}
Let us now turn to the discussion of the temperature dependence of the spectral 
functions calculated with CFS method. Here, an additional problem arises, because
the temperature scale in NRG is defined by the length $N$ of the chain.
More precisely, a
given chain length $N$ corresponds to a temperature $T_N\sim\Lambda^{-(N-1)/2}$.
Thus, a continuous variation of $T$ is not possible, only a discrete set of
logarithmically varying temperatures is accessible. More problematic, however,
is that due to the termination of the NRG iterations at chain length $N$ no
information are available for {\em excitation energies} smaller than $T_N$, too.\cite{BullaCostiVollhardt01}
This means, that the part of the spectrum with $|\omega|<T_N$ is inaccurate in 
the sense that excitations on these scales cannot be taken into account properly.
Furthermore, the limited information available for this energy range stems from
all NRG iterations, and hence a Gaussian broadening (\ref{eqn:broadening}) 
suitable for excitations collected from a single iteration cannot be used
here. Instead, a Lorentzian broadening\cite{BullaCostiVollhardt01}
\begin{equation}\label{eqn:lorentian_broadening}
\delta(\omega-\omega_n)\to\frac{1}{2\pi}\frac{\tilde{b}}{(\omega-\omega_n)^2+\tilde{b}^2}
\end{equation}
is used for $|\omega|<\alpha T_N$, where $\alpha$ sets the energy scale down
to which we trust the NRG results.\footnote{For our calculations, we choose $\tilde{b}=0.8$ and $\alpha=2$.}
Moreover, as discussed in detail in the appendix, 
the CFS still ensures a complete basis set and hence an exact norm for
the spectral functions, but quantities obtained from spectral averages like occupancies
must be expected to be less accurate than for $T=0$.

 \begin{figure}[tb]
   \centering
   \includegraphics[width=80mm,clip]{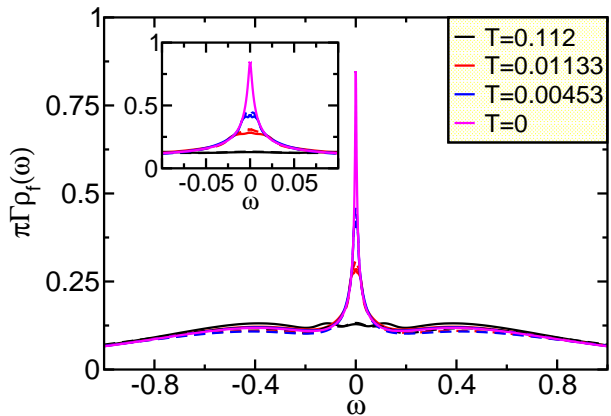}

   \caption{(color online) Spectral function for temperatures $T=0$,
     $0.00453\approx T_K$, $0.01133$ and $0.112\gg T_K$ calculated
     with CFS. The full curves result from calculations with
     $N_S=500$, the dashed ones from calculations with $N_S=50$.
Model and NRG parameters are otherwise the same as in Fig.~\ref{fig:2}.}
   \label{fig:6}
 \end{figure}

The development of the spectral function for $\Gamma/D=\pi V^2\rho/D=0.1$, 
$\epsilon_f=-U/2=-0.5D$ as function of temperature 
can be found in Fig.~\ref{fig:6}.  Even though we restrict the results
presented in this section to the particle-hole symmetric limit, all
observations also hold for the asymmetric case with and without
magnetic field. As expected, with increasing temperature the ASR is
reduced, for $T\approx T_K$ 
to roughly half its original height. For $T\gg T_K$ it has completely vanished,
leaving only the Hubbard peaks in the spectrum. For all spectra we find
the norm to be exactly one as before and, due to particle-hole symmetry, the
occupancy has to be one-half. The full curves in Fig.~\ref{fig:6} are results
of NRG calculations with $N_S=500$ states, the dashed ones with
$N_S=50$ states. As before, we can notice mild deviations of the
spectra in the region of the Hubbard peaks and at very low energies,
but the overall agreement of the spectra is quite good, thus again
underlining the previous claim that spectra calculated with the CFS
are rather insensitive to the number $N_S$ of states kept in each NRG
iteration. Note that the deviations for $\omega\to0$ can be easily
accounted for by the fact that with varying $N_S$ the distribution of
excitation energies 
in the NRG will differ, thus yielding a slightly different distribution
of spectral weight due to the broadening (\ref{eqn:lorentian_broadening}).

\begin{figure}[htb]
   \centering
   \includegraphics[width=80mm,clip]{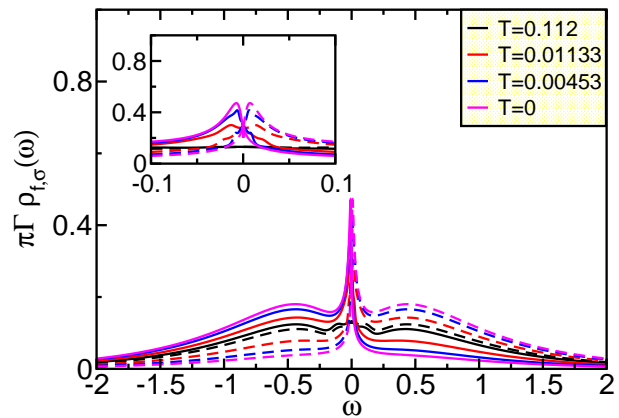}
   
   \caption{(color online) Spectral function for temperatures $T=0$,
     $0.00453\approx T_K$, $0.01133$ and $0.112\gg T_K$ and finite magnetic
     field $H=0.005$ calculated
     with CFS. The full curves show the spectral function for spin up, the
     dashed curves spin down. Model and NRG parameters are otherwise the same as in Fig.~\ref{fig:3}. The values for norm and occupancies are listed in
Tab.~\ref{tab:symm_with_b_finite_T}.}
   \label{fig:7}
 \end{figure}

The calculations for finite magnetic field $H=0.005$ are collected in Fig.~\ref{fig:7}.
Here we only present results for $N_S=50$, because as noted before 
the spectra for $N_S=500$ do not differ significantly. As in Fig.~\ref{fig:3}
we plotted the spectra for the majority spin as full curves, the ones for the
minority spin as dashed curves. Again, the norm is exactly one in all cases. The
occupancies obtained from integrating the spectra multiplied with the Fermi
function at the appropriate temperature are listed in
\begin{table}[htb]
\setlength{\tabcolsep}{2mm}
\centering
\begin{tabular}{rrrrrrr}
\multicolumn{1}{c}{$N_S$} 
 & \multicolumn{1}{c}{$T$} 
 & \multicolumn{1}{c}{Norm} 
 & \multicolumn{2}{c}{$\langle n_\sigma\rangle_{GF}$} 
 & \multicolumn{2}{c}{$\langle n_\sigma\rangle_{NRG}$}\\
 &
 &
 & \multicolumn{1}{c}{$\sigma=\uparrow$} 
 & \multicolumn{1}{c}{$\sigma=\downarrow$} 
 & \multicolumn{1}{c}{$\sigma=\uparrow$} 
 & \multicolumn{1}{c}{$\sigma=\downarrow$} \\
\hline 
500 & $0$ & 1 & 0.806 & 0.194 & 0.806 & 0.194\\
50 & $0$ & 1 & 0.807 & 0.193 & 0.807 & 0.193\\
500 & $0.00453$ & 1 & 0.723 & 0.277 & 0.723 & 0.277\\
50 & $0.00453$ & 1 & 0.729 & 0.271 & 0.729 & 0.271\\
500 & $0.01133$ & 1 & 0.629 & 0.371 & 0.629 & 0.371\\
50 & $0.01133$ & 1 & 0.635 & 0.365 & 0.635 & 0.365\\
500 & $0.112$ & 1 & 0.518 & 0.482 & 0.518 & 0.482\\
50 & $0.112$ & 1 & 0.519 & 0.481 & 0.519 & 0.481\\
\hline
\end{tabular}
\caption{Norm and occupancies for the calculations with finite magnetic field
and $T>0$ in Fig.~\ref{fig:7}. In addition the occupancies for $N_S=500$ are
included in the table. The last columns contain the occupancies obtained from
the thermodynamic expectation values.\label{tab:symm_with_b_finite_T}}
\end{table}
Tab.~\ref{tab:symm_with_b_finite_T} for $N_S=500$ and $N_S=50$. 
To avoid additional inaccuracies introduced by the broadening
(\ref{eqn:lorentian_broadening}) we have listed in Tab.~\ref{tab:symm_with_b_finite_T} 
the sum of the weights of the raw NRG spectra multiplied with the
Fermi function. As for $T=0$, the deviations to the thermodynamic
values stay always less than $0.01$\%. 
Numerical integration of the broadened spectra, on the other hand, will lead
to stronger deviations,
in particular for higher temperatures. For example, we find
$\langle n_\uparrow\rangle_{GF}=0.516$ for $T=0.112$ and $N_S=500$, i.e.\ an
error in the percent range. The value of  $\int\rho(\w)f(\w)$ with
$\rho(\omega)$ smoothened by the 
described combination of Gaussian and Lorentzian broadening
deviates from the sum over the spectral weights of the poles of the discrete spectrum
times the Fermi function at the pole energy, since the Lorentzian part extends 
in an uncontrolled manner to positive and negative frequencies.  This
systematic error, however, is still small even in these extreme cases.

\section{Conclusion}

We presented a novel approach for the calculation of local Green's functions
for quantum impurity systems using Wilson's numerical
renormalization group. Since it is based on the usage of a {\em
  complete basis set} for the Wilson NRG chain, recently introduced in
the context of the time-dependent
NRG,\cite{AndersSchiller2005,AndersSchiller2006} quantities written as
spectral sum rules are 
{\em always exactly} fulfilled, regardless of the size $N_S$ of the
Hilbert space kept in each NRG iteration. Moreover, spectral averages
like occupation numbers etc.\ are at least reproduced with unprecedented
accuracy. We
have demonstrated the quality of the approach by presenting results for the 
fermionic single-particle spectral function $\rho_f(\w)$ of the single
impurity Anderson model. We have shown explicitly the validity of our
claim, that the resulting spectral functions are very insensitive to
the number of NRG states kept in the iteration. This is a consequence
of using a complete basis set in the derivation of the Lehmann
representation. Equations (\ref{eqn:green-i})-(\ref{eqn:green-iii})
precisely account for which excitation contributes at which energy
shell $m$, thus manifestly solving the double counting problem of
excitations responsible for the violation of the sum rules in all
previous NRG spectral function approaches: The conventionally employed
``patching'' algorithms thus have become obsolete. Also, no even-odd
oscillations can be observed. We believe that our  novel approach
provides the most accurate  representation of local spectral functions
for a given NRG input. It can be used in quantum impurity systems, at
finite temperature as well as $T\to 0$. It is particular suitable for
symmetry broken phases, for instance in an external magnetic field.

This significant improvement can have a major impact on the usage of
the NRG as impurity solver for the dynamical mean field
theory (DMFT).\cite{Pruschke95,Georges96} In particular, for the description
of symmetry broken phases such as the ferromagnetic Hubbard
model\cite{ZIZLERXXX}  as well as orbital order in the two-band
Hubbard model, the discrepancy between the local NRG order parameter
and the one obtained from the spectral function often yields
instabilities in iterating the DMFT equations. Moreover, 
very accurate spectral functions  are needed  for the application of
the NRG to cluster   DMFT approaches,\cite{MaierJarrellPruschkeHettler2005} or periodic
Anderson models which include crystal field effects\cite{AndersPruschke2006}.
Up to now such an accuracy could only be achieved with immense CPU
time.\cite{PruschkeBullaTwoBandHubbard2005}

\begin{acknowledgments}
We acknowledge stimulating discussions with A.\ \mbox{Weichselbaum}
which helped us clarifying that CFS can be applied to the evaluation
of correlation functions rigorously such that inter-shell contributions
are absent. After completion of this work we learned that Weichselbaum
and von Delft independently  had followed up on the same
idea\cite{WeichselbaumDelft2006} with a slightly different focus
and a more generalized treatment of the density matrix.
We further thank T.\ A.\  Costi for stimulating discussions, as well as A.\ Schiller for
collaborations on related work. 

This research was supported by the DFG through collaborative research
grant
SFB 602 (RP and TP) and project AN 275/5-1 (FBA). We acknowledge
supercomputer support by the Gesellschaft f\"ur wissenschaftliche
Datenverarbeitung G\"ottingen and the Norddeutsche Verbund f\"ur
Hoch- und H\"ochstleistungsrechnen under project nip00015 (RP and TP), as well as the NIC, Forschungszentrum
J\"ulich under project no.\ HHB000 (FBA). 
\end{acknowledgments}

\appendix

\section{Proof of the Spectral Sum Rule}
\label{sec:a1}

By inserting the analytical expression (\ref{eqn:green-tot}) for the
discrete spectral function into the integral (\ref{eqn:contour})
\begin{equation}
  \label{eq:a1}
C = \oint \frac{dz}{2\pi i} G_{A,B}(z) = \sum_{\alpha=i,ii,iii} \oint
\frac{dz}{2\pi i} G^{\alpha}_{A,B}(z)  
\end{equation}
yields the following three contributions
\begin{eqnarray}
\label{eqn:a2-contour}
  C &=& 
\sum_{l,l'}
A_{l,l'}(N) B_{l',l}(N)
\left(e^{-\beta E^N_l}-s \,e^{-\beta E^N_{l'}}\right)
\\
&& +
\sum_{m=m_{min}}^{N-1}
 \sum_{l}
\sum_{k,k'} B_{l,k'}(m)\rho^\textrm{red}_{k',k}(m) A_{k,l}(m)
\non
&&
-s\sum_{m=m_{min}}^{N-1}
 \sum_{l}
 \sum_{k,k'} A_{l,k'}(m) \rho^\textrm{red}_{k',k}(m)B_{k,l}(m)
\nonumber
\punkt
\end{eqnarray}
By formally defining the ``reduced density matrix'' for the last
Wilson shell as 
\begin{equation}
  \label{eq:reduced-rho-N}
  \rho^{red}_{kk'}(N) = \delta_{k,k'} \frac{e^{-\beta E_k^N}}{Z}  
\komma
\end{equation}
the first term in  (\ref{eqn:a2-contour}) can be included into the
second and third. Then the second reads
\begin{eqnarray}
  \label{eq:a4}
  C_2 &=&
\sum_{m=m_{min}}^{N}
 \sum_{l'}
\sum_{k,k'} B_{l',k'}(m)\rho^\textrm{red}_{k',k}(m) A_{k,l'}(m)  
\non
%%----------------------------------------------
&=&
\sum_{m=m_{min}}^{N}
 \sum_{l'e''} \sum_{k,e}\sum_{k',e'} 
\bra{l',e'';m} B\ket{k',e';m}
\non
%%&& 
&&\times \bra{k',e';m}\hat \rho \ket{k,e;m}
%\non
%%&& 
\bra{k,e;m} A \ket{l',e'';m}
\end{eqnarray}
In this step, we used the definition of the reduced density matrix and
made use of the fact that the local operators $A$ and $B$ are
independent of the environment variables $e,e',e''$. Due to the form
of the density operator, $(1^+_m +1_m^-)\rho(1^+_m +1_m^-) = 1^+_m\rho
1^+_m$ holds for $m<N$. For $N=m$, $1^-_N$ spans the complete Fock
space and   the projection $1_m^-\rho 1_m^-$ contains all
contributions at $m=N$. Therefore, the contribution  $C_2$ yields
\begin{eqnarray}
  \label{eq:a5}
  C_2 =
\sum_{m=m_{min}}^{N}
 \sum_{l'e''} 
\bra{l',e'';m} B \rho A  \ket{l',e'';m}
\non
= 
 \Tr{\rho A B}
\punkt
\end{eqnarray}
We can perform the same calculation for the third term in
(\ref{eqn:a2-contour}) and the contribution stemming from the $(-s)$
part of the first term at $m=N$ to derive
\begin{eqnarray}
  C_3 = -s \Tr{\rho  B A}
\punkt
\end{eqnarray}
To this end, the total contribution is given by
\begin{equation}
  C =  \Tr{\rho A B}-s \Tr{\rho  B A} =  \Tr{\rho [A,B]_{s}}
\end{equation}
We thus have proven that our  Green's function fulfills the spectral sum
rule {\em exactly}. Thus, any deviations for this sum rule in the
broaden spectral function stems solemnly from the accuracy of the
numerical $\w$ integration.

\section{Occupation sum rule}
\label{app:occ-sum-rule}

We now want to discuss the accuracy of our approach for the
generalized occupancy $n_{B,A}$ defined as
\begin{equation}
  \label{eq:aa1}
n_{B,A} = \oint \frac{dz}{2\pi i} f_s(z) G_{A,B}(z)\;,
\end{equation}
where $f_s(z)= [\exp(\beta z) -s]^{-1}$. As always throughout the paper,
this definition comprises
bosonic and fermionic Green's functions. 
Specializing  to the local spin-dependent fermionic spectral
function, i.e.\ $A=f_\sigma$ and $B=f^\dagger_\sigma$, yields
the occupational sum rule (\ref{equ:occupation-sum-rule}) as presented
in section \ref{sec:occ}.

Inserting the Green's function
(\ref{eqn:green-tot}) into the expression (\ref{eq:aa1}) and performing the contour
integration yields the following three contributions:
\begin{eqnarray}
  n_{B,A} &= &\sum_{l,l'} \frac{e^{-\beta E_l}}{Z} B_{l,l'}(N) A_{l',l}(N)
\\
&& +
\sum_{m=m_{min}}^{N-1}
 \sum_{l}
\sum_{k,k'} 
\frac{B_{l,k'}(m)\rho^\textrm{red}_{k',k}(m) A_{k,l}(m)}{e^{\beta(E_{l}^m-E_{k}^m)} -s}
\non
&&
+
\sum_{m=m_{min}}^{N-1}
 \sum_{l}
 \sum_{k,k'} 
\frac{A_{l,k'}(m) \rho^\textrm{red}_{k',k}(m)B_{k,l}(m)}{1-se^{\beta(E_{k}^m-E_{l}^m)}}
\nonumber
\punkt
\end{eqnarray}
Adding and subtracting
$$
\sum_{m=m_{min}}^{N-1}
 \sum_{l}
\sum_{k,k'} 
\frac{A_{l,k'}(m)\rho^\textrm{red}_{k',k}(m) B_{k,l}(m)}{e^{\beta(E_{l}^m-E_{k}^m)} -s}
$$
leads to
\begin{eqnarray}
  n_{B,A} &=&\sum_{l,l'} \frac{e^{-\beta E_l}}{Z} B_{l,l'}(N) A_{l',l}(N)
\\
&&
+
\sum_{m=m_{min}}^{N-1}
 \sum_{l}
 \sum_{k,k'} A_{l,k'}(m) \rho^\textrm{red}_{k',k}(m)B_{k,l}(m)
\non
&& +\delta n_{B,A}\non
&=& \Tr{\rho B A}+\delta n_{B,A}\;,
\nonumber
\end{eqnarray}
using the same identities as in appendix \ref{sec:a1}. Finally, the
error $\delta n_{B,A}$ of the occupation sum rule is given by
\begin{eqnarray}
  \delta n_{B,A} &=&
\sum_{m=m_{min}}^{N-1}
\sum_{l,k,k'} \rho^\textrm{red}_{k',k}(m)
\\
&&
\times
\frac{B_{l,k'}(m) A_{k,l}(m)-A_{l,k'}(m)B_{k,l}(m)}{e^{\beta(E_{l}^m-E_{k}^m)} -s}\;.\nonumber 
\end{eqnarray}
To estimate the size of this error, let us note that the excitation energies
$E_l^m-E_k^m$ appearing in $\delta n_{B,A}$ are positive, because
$l$ labels a discarded and $k$ a kept state. Excitations between two
discarded states never contribute as well as excitations between to
kept states. The smallest energy difference  possible is given by the
energy of the first discarded  minus the energy of the last kept
state. Even though this energy difference might be small, the
smallness of the matrix element of the reduced density matrix
suppresses this contribution. We are thus left with energy differences
which are of the order $D_m$ for $m<N$. The density matrix $\hat\rho$,
on the other hand, is restricted to the last iteration, i.e., to be
consistent with the fundamental assumption of the NRG,
$\exp[\beta(E_l^m-E_k^m)] \gg1$ for $m<N$.  Thus, the denominator in
$\delta n_{B,A}$ becomes exponentially large and 
hence $\delta n_{B,A}$ exponentially suppressed with decreasing temperature.
Note that this also means, that for $T\to0$ the occupation sum rule will be
obeyed exactly, too. For finite $T$ we must expect, and indeed find
(see section \ref{sec:finite_T}), deviations from this sum rule with
increasing temperature. Our numerical results show, however, that even
for high temperatures these deviations stay below the per mille level.

\section{Imaginary time Green's functions}

The textbook definition of the imaginary time Green's function 
is given by
\begin{eqnarray}
\label{eqn:6a}
  G_{A,B}(\tau) &=& -\Tr{\hat{\rho} T\left(A(\tau) B\right)}
\nonumber \\
 & =& \left\{
    \begin{array}{ccc}
      -\Tr{\hat{\rho} e^{\tau H} A e^{-\tau H} B } & ; &\tau >0 \\
      -s\Tr{\hat{\rho} B e^{\tau H} A e^{-\tau H}  } &; &  \tau <0
    \end{array}
\right.
\end{eqnarray}
where again $s=1$ for Bosonic operators $A,B$ and $s=-1$ for
Fermionic operators. It is straight forward  to show that the
analytic continuation $G_{A,B} (z)$ of the {\em exact} Matsubara
Green's function 
$G_{A,B} (i\omega)$ 
\begin{eqnarray}
  G_{A,B} (i\omega) &=& \int_0^\beta d\tau e^{i\omega \tau}  G_{A,B} (\tau) \;,
\komma
\end{eqnarray}
where $\omega = \pi(2n+1)/\beta$ for Fermionic and $\omega = 2\pi
n/\beta$ for Bosonic Green's functions, is identical to the Laplace
transformed {\em exact} retarded  Green's function
(\ref{eqn:lehmann-representation}) with $z=i\omega$. 
We can perform the same steps for $G_{A,B}(\tau)$ as in section
\ref{sec:II.C}. In this case, however, the terms
$-s[z +E^m_{l}-E^m_{k}]^{-1}$ in $G^{ii}(z)$ and 
$[z  +E^m_{k}-E^m_{l}]^{-1}$ in  $G^{iii}(z)$  must be replaced by
$$
\frac{\exp(\beta(E^m_{k} - E^m_l)-s}{i\omega +E^m_{l}-E^m_{k}}
$$ 
and 
$$
\frac{1-s\exp[\beta(E^m_{k} - E^m_l)]}{i\omega  +E^m_{k}-E^m_{l}}\;.
$$
At first sight, the two Green's
functions are not equal in our approach. However, as discussed in  appendix
\ref{app:occ-sum-rule}, the Boltzmann
factors $\exp[\beta(E^m_{k} - E^m_l)]$ can be neglected for $m<N$. This is
justified by the NRG assumption that the density operator is well
approximated by its contribution of the last Wilson shell, i.e. Eq.\
(\ref{eqn:rho-nrg}). Therefore,  both Green's functions are identical
within the NRG approximation.

%%%%%%%%%%%%%%%%%%%%%%%%%%%%%%%%%%%%%%%%%%%%%%%%%%%%%%%%%%%%%%%%%%%%%
%%\bibliography{references}

\end{document}